\documentclass{aa}  

\usepackage{graphicx}
\usepackage{xcolor}
\usepackage{txfonts}
\usepackage{subcaption}         % necessary for continued figures, example in section 3
\usepackage{lscape}             % to rotate a single page table, example in appendix.
\usepackage{placeins}           % useful with \FloatBarrier, to keep 
\usepackage{etoolbox}
\usepackage{multirow}
\usepackage{cleveref}
\newcommand{\cloudy}{\textsc{cloudy}}
\newcommand{\kms}{km s$^{-1}$}
\newcommand{\cmN}{cm$^{-2}$}

\newcommand{\lam}{$\lambda$}

\newcommand{\lya}{\mbox{Ly$\alpha$}}

\newcommand{\hi}{\mbox{H\,{\sc i}}}

\newcommand{\civ}{\mbox{C\,{\sc iv}}}

\newcommand{\siiv}{\mbox{Si\,{\sc iv}}}

\newcommand{\nv}{\mbox{N\,{\sc v}}}

\newcommand{\mgii}{\mbox{Mg\,{\sc ii}}}
\newcommand{\aliii}{\mbox{Al\,{\sc iii}}}

\newcommand{\target}{J004613+010425}                                
\begin{document}

%%%%%%%%%%%%%%%%%%%%%%%%%%%%%%%%%%%%%%%%
% if you use custom commands in your title,
% ensure to check your title when submitting!
%%%%%%%%%%%%%%%%%%%%%%%%%%%%%%%%%%%%%%%%
   \title{Resolving the Complex Line-Locking Structure in Quasar J004613+010425 with VLT/X-shooter Spectroscopy}

%   \subtitle{Subtitle}

%%%%%%%%%%%%%%%%%%%%%%%%%%%%%%%%%%%%%%%%
% Please separate each author with the \and command
%
% Use the \corrauth to provide the corresponding
% author address. It will be automatically inserted as 
% footnote in the PDF output.
%
% Please DO NOT include ORCIDs next to author names.
% Instead, please provide an active address for each coauthor:
% it will be automatically extracted by EDPS editorial system, 
% and co-authors will be be able to authenticate their ORCID.
%
% Only authenticated ORCIDs will be taken into account.
% ORCIDs included here will be removed.
%%%%%%%%%%%%%%%%%%%%%%%%%%%%%%%%%%%%%%%%

   \author{Chen Chen\inst{1}\corrauth{chenchen21@zcst.edu.cn}        % use \corrauth for the corresponding author
        \and Wei-Jian Lu\inst{2}\email{william_lo@qq.com}
        \and Ying-Ru Lin\inst{3}\email{1015000842@qq.com}
        \and Bo Ma\inst{4,5}\corrauth{mabo8@mail.sysu.edu.cn}
        }

   \institute{Zhuhai College of Science and Technology, Zhuhai 519000, China
   \and School of Information Engineering, Baise University, Baise 533000, China
   \and College of Chemical and Environmental Engineering, Baise University, Baise 533000, China
   \and CSST Science Center in the Great Bay Area, Sun Yat-sen University, Zhuhai 519000, China
   \and School of Physics $\&$ Astronomy, Sun Yat-Sen University, Zhuhai 519000, China}

   \date{Received }

% \abstract{}{}{}{}{}
% 5 {} token are mandatory
 
  \abstract
   {Line-locking (LL) signatures in quasar absorption-line systems provide direct observational evidence for radiatively driven outflows. We present a detailed investigation of LL phenomena in the quasar SDSS~J004613.54+010425.7 ($z_{\rm em} = 2.150$) using medium-resolution VLT/X-shooter spectra, as part of the TOLL project. Compared with previous studies based on Sloan Digital Sky Survey spectroscopy, the X-shooter data resolve the absorption complex into twelve narrow absorption line (NAL) components spanning $\sim$$2700$ \kms\ in the \civ, \nv, \siiv, and \lya\ transitions, a factor of $\sim$2.4 more components than identified from lower resolution spectra. We detect fourteen distinct LL pairs across three ionic species, forming one of the largest line-locking webs known to date: three \siiv\ pairs connecting components (2, 10), (4, 11), and (5, 12); six \nv\ pairs organized into four groups; and five \civ\ pairs organized into three groups. Monte Carlo simulations show that the probability of this twelve-component LL web arising by chance is $< 10^{-5}$. Photoionization modeling with \cloudy\ yields ionization parameters of $\log U \approx -2.4$ to $-0.6$ and total hydrogen column densities of $\log N({\rm H}) \approx 18.4$--$20.6$ for the twelve components. A comparison of the properties of shadowed and shadowing clouds shows that ion column densities and ionization parameters typically agree within a factor of $\sim$3, broadly supporting the standard radiative acceleration picture, although several outliers suggest that multi-ion locking and cloud-cloud interactions may also regulate LL stability. These results demonstrate that medium- and high-resolution spectroscopy is essential for resolving the full complexity of line-locked quasar outflows.}


   \maketitle
   \nolinenumbers

%%%%%%%%%%%%%%%%%%%%%%%%%%%%%%%%%%%%%%%%%%%%%%%%%%%%%%%%%%%%%%
\section{Introduction}

Quasars, powered by accreting supermassive black holes, rank among the most energetic objects in the Universe. Quasar outflows, observed as blueshifted absorption features in rest-frame ultraviolet (UV) spectra, are an important manifestation of the interaction between accreting supermassive black holes and their host galaxies \citep{Silk98, Kauffmann00, Berti08, Costa14, Barai16, Angl17a, Angl17, Barai19, Bustamante19, He22, Chen22, Bluck23, Ayubinia23, Naddaf23, Hall24, He24, Bollati24}. These outflows span a wide range of velocity widths and ionization states, giving rise to three main classes that we list here in order of decreasing velocity width: broad absorption lines (BALs; absorption troughs wider than $\sim2000$~\kms), mini-BALs ($\sim500$--$2000$~\kms), and narrow absorption lines (NALs; typically $\lesssim500$~\kms). Their kinematic and ionization structures provide important constraints on the physical conditions, acceleration mechanisms, and feedback effects of active galactic nuclei (AGN)-driven winds.

Among the observed signatures of quasar outflows, line-locking (LL) is particularly noteworthy. In line-locked systems, the velocity separation between two absorption components closely matches the laboratory doublet splitting of a resonance doublet transition, such as \civ\ (498 \kms) or \nv\ (964 \kms). This phenomenon has long been regarded as evidence for radiative acceleration in quasar winds \citep{Milne26, Scargle73, Burbidge75, Goldreich76, Foltz87, Braun89}, and its origin in radiatively accelerated outflows has been investigated theoretically by, e.g., \citet{Lewis23}. In the canonical picture, a slower shadowing cloud lies between the continuum source and a faster shadowed cloud. The resulting reduction in radiative force acting on specific transitions can establish a preferred velocity separation between the two components. As a result, line-locking provides a direct observational probe of the role of radiation pressure in accelerating and shaping quasar outflows. Theoretical studies have explored the formation and long-term evolution of line-locked absorbers arising from resonance-line interactions in active galaxies.

A number of observational studies have reported line-locking signatures in BAL quasars using Sloan Digital Sky Survey (SDSS) spectra, which have a spectral resolution of $R\sim2000$. \citet{Lu25} analyzed seven epochs of SDSS observations of J004613.54+010425.7 (hereafter \target) and identified five \siiv\ NAL doublets within BAL system B, including several line-locked pairs with velocity separations consistent with the \civ\ and \aliii\ doublet splittings. 
Similarly, \citet{Lin20a} investigated two epochs of SDSS spectra of J115122.14+020426.3 (hereafter J1151+0204) and reported a ``quintuple'' line-locked structure composed of four \civ\ NAL doublets in BAL system C. 
These studies demonstrate that line-locking can be common in complex BAL outflows. However, the limited spectral resolution of SDSS data prevents narrow absorption structures from being fully resolved. This can introduce significant uncertainties into measurements of NAL quantities that are essential for robust line-locking analyses, including velocities, component multiplicities, and equivalent widths (EWs).

Medium- and high-resolution spectroscopy offers a direct way to overcome these limitations. Instruments such as the Very Large Telescope (VLT)/X-shooter and VLT/UVES achieve spectral resolutions of $R\gtrsim5000$, allowing closely spaced NAL components to be resolved and their properties to be measured with substantially improved accuracy \citep{Arav18, Xu19, Miller20}. Such observations are essential for testing whether apparently simple line-locked configurations identified in low-resolution spectra in fact comprise more complex absorption structures. 

This study is part of the TOLL project \citep{Chen24, Chen25}, which is dedicated to the investigation of line-locking phenomena using medium- and high-resolution quasar spectra. Hereafter, we refer to \citet{Chen24} and \citet{Chen25} as Paper~I and Paper~II, respectively. In this paper, we present medium-resolution VLT/X-shooter observations of \target\ and compare them directly with previously published SDSS spectra. Our primary goal is to investigate how spectral resolution affects the identification and interpretation of line-locking systems and to quantify the biases introduced by low-resolution spectroscopy. By resolving the detailed substructure of the absorption complexes, we reassess the morphology of the reported line-locked systems and examine the extent to which medium- and high-resolution observations modify the inferred outflow properties. 
We note, however, that narrow metal absorption lines are generally only partially resolved even with X-shooter. Thus, while the present data represent a major improvement over the SDSS spectra ($R\sim2000$), they do not provide a fully resolved view of the NAL components; higher-resolution observations (e.g., with VLT/UVES at $R\sim10^{4}$--$10^{5}$) would be required to fully resolve them.

This paper is organized as follows. Section~2 describes the observations and data analysis procedures. Section~3 presents the results. In Section~4, we discuss the implications of our findings for the interpretation of line-locking and quasar outflow physics. Section~5 summarizes our conclusions. Throughout this paper, we adopt a cosmological model with $H_0 = 71$ \kms\ Mpc$^{-1}$, $\Omega_M = 0.27$, and $\Omega_\Lambda = 0.73$.

\section{Quasar spectra and data analysis}

 \subsection{Medium-Resolution Spectra}

We retrieved archival medium-resolution spectra of quasar \target\ from the European Southern Observatory (ESO) Science Archive Facility. These datasets provide an excellent opportunity to investigate line-locking structures. The key observational parameters are summarized in \Cref{tab:data}. 

\target\ was observed with the X-shooter spectrograph on the Very Large Telescope (VLT). The available data were obtained through two ESO programs. Program 090.B-0577 (PI, principal investigator: B. Punsly; 2012) consists of single exposures of 2800--3000 s and covers 298.9--2479 nm. Program 091.B-0324 (PI: C. Benn; 2013) comprises single exposures of 3440--3600 s, covering 298.9--2479 nm. The slit widths were $1.6''$, $1.5''$, and $1.2''$ for the UVB (ultraviolet), VIS (visible), and NIR (near-infrared) arms for program 090.B-0577, and $1.0''$, $0.9''$, and $0.9''$ for program 091.B-0324, respectively. Since the resolving power is approximately constant within each X-shooter arm, we quote the resolving powers measured from the calibration data as provided by the ESO Phase~3 products: $R_{\rm UVB}\approx3251$, $R_{\rm VIS}\approx5040$, and $R_{\rm NIR}\approx4287$ for 090.B-0577, and $R_{\rm UVB}\approx5453$, $R_{\rm VIS}\approx8935$, and $R_{\rm NIR}\approx5573$ for 091.B-0324. The total on-source exposure times are 3000/2800/3000~s (UVB/VIS/NIR) for 090.B-0577 and 10800/10320/10800~s for 091.B-0324. We find no evidence for significant absorption-line variability among the different observations. All spectra were therefore coadded to produce a single spectrum with higher signal-to-noise ratio (S/N) for subsequent analysis. The coadded spectrum has a pixel size of 0.2~\AA\ pix$^{-1}$ in the UVB arm and a typical S/N of $\sim$70 per pixel in the continuum regions redward of the \lya\ emission line. Because the coadd is dominated by the higher-resolution and higher-S/N 091.B-0324 exposures, we adopt an effective resolving power of $R_{\rm eff}\approx5400$, corresponding to an instrumental velocity dispersion of $\sigma_{\rm inst}\approx23.6$~\kms, for the combined UVB spectrum; this value is used in the line-profile fits (Section~2.3).

\begin{table*}[ht!]
\caption{Quasar data including the name, emission-line redshift, telescope-instrument used for observations, program ID, Observation Date (year-mm-dd) and MJD (Modified Julian Date), wavelength coverage, and spectral resolution $R=\lambda /\Delta\lambda$.}
\label{tab:data}
\centering
\renewcommand{\arraystretch}{1.5}
\begin{tabular}{ccccccc}
\hline\hline
Quasar & $z_{\rm{em}}$ & Instrument & Program ID & Obs. Date (MJD) & $\lambda$ (\r{A}) & $R$\\
\hline
\target\ & 2.150 & VLT/X-shooter & 090.B-0577 & 2012-11-28 (56259) & 2989-24790 & 3251-5040\\
  &  &  & 091.B-0324 & 2013-07-22 to 2013-08-02 (56495-56506) & 2989-24790 & 5453-8935\\
\hline
\end{tabular}
\end{table*}

\subsection{Data Reduction and Continuum Normalization}

All spectral datasets were obtained directly from the ESO Science Archive Facility as pre-reduced products, which had been processed through the standard ESO X-shooter pipelines to perform bias subtraction, flat-fielding, wavelength calibration, and flux calibration. 
To place all spectra on a consistent continuum scale, we carried out an additional continuum-normalization procedure. Spectral regions free of strong emission and absorption features were identified and fitted iteratively with power-law continuum functions. 
The resulting continuum models were combined with Gaussian representations of the prominent emission lines (e.g., \civ\ and \siiv) to construct a pseudo-continuum for each spectrum.
The observed spectra were then divided by their corresponding pseudo-continua, yielding normalized spectra with a continuum level of unity. This normalization enables direct comparisons of absorption features across different wavelength regions and observing epochs.

\begin{figure*}
\centering
\includegraphics[width=0.85\textwidth]{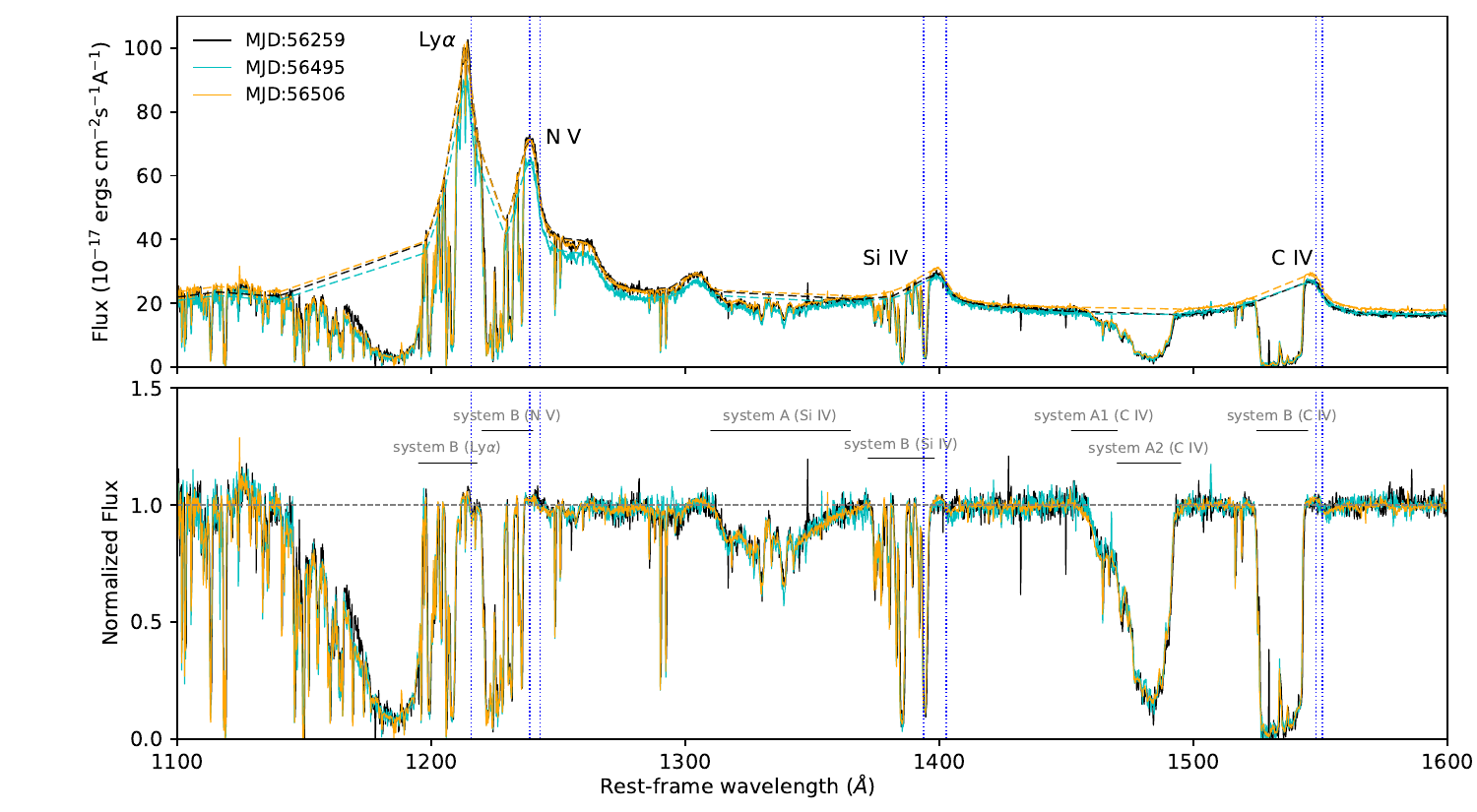}
\caption{VLT/X-shooter spectra of \target\ at the rest-frame wavelengths (defined by the emission redshift $z=2.150$). Top panel: the three-epoch spectra (solid curves) and the corresponding pseudocontinuum fits (dashed curves). Bottom panel: the normalized spectra. Black horizontal solid bars indicate two systems.}\label{fig:J0046_total}
\end{figure*}

\subsection{Absorption Line Identification and Fitting}
Absorption systems were identified and characterized following the procedures described in \citet{chen18, Chen19, chen20, Chen24, Chen25}. Our analysis focuses on associated narrow absorption-line (NAL) systems. We define associated absorbers as those lying within $|\Delta v| \lesssim 8000$~\kms\ of the quasar emission redshift \citep{Chen20a}. Within this velocity window, we identified all narrow absorption lines (system B in \Cref{fig:J0046_total}) that satisfy this criterion. The emission redshift $z_{\rm em}=2.150$ is taken from the improved SDSS quasar redshift catalog of \citet{Hewett10}, which was also adopted in the previous study of this object \citep{Lu25}. In most cases, we use the \civ\ \lam1548,1551 doublet as the primary tracer because it is typically strong in our spectra, lies outside the Ly$\alpha$ forest, and has a high signal-to-noise ratio. For \target, however, the \civ\ lines are nearly saturated and heavily blended, so we adopt the \siiv\ doublet as the primary tracer to identify individual absorption components.

We first identified velocity components within the associated narrow absorption-line (NAL) complexes using the \siiv\ doublet. For each component, we then searched for corresponding transitions of \civ\ \lam1548,1551, \nv\ \lam1239,1243, and \lya\ \lam1216 at the same redshift. Any additional components revealed through these transitions were subsequently cross-checked against the other ionic species. This iterative procedure was repeated until all absorption features associated with each system were consistently identified and cataloged.

\begin{figure*}[ht!]
\centering
\includegraphics[width=0.8\textwidth]{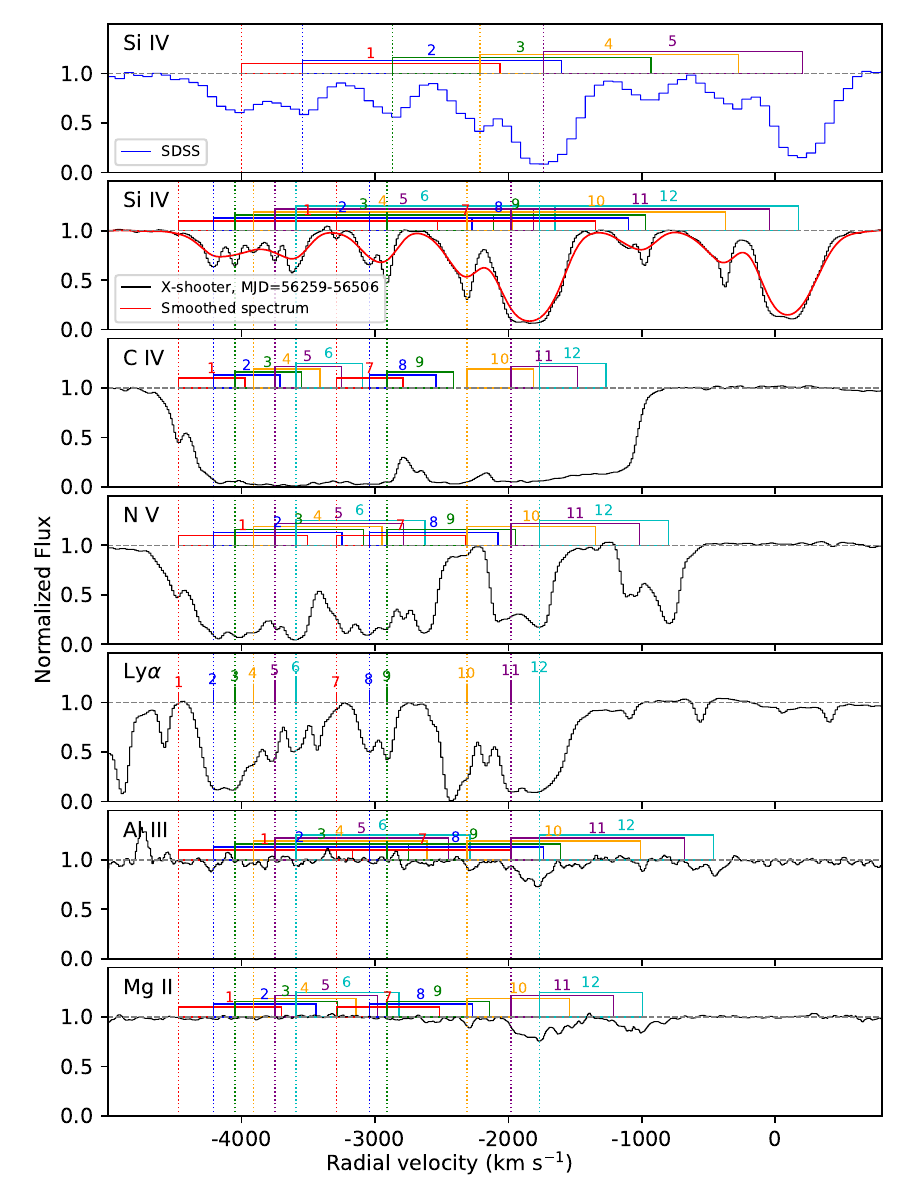}
\caption{Rest-frame normalized flux spectra of \target\ ($z=2.150$) as a function of radial velocity offset from systemic redshift, covering \siiv, \civ, \nv, \lya, \aliii, and \mgii\ from top to bottom. The first panel shows the SDSS \siiv\ spectrum for reference; all other panels display X-shooter spectra (MJD=56259--56506). In the second panel, the red smoothed X-shooter spectrum matches the overall absorption structure seen in the SDSS \siiv\ data. Colored vertical bands indicate the velocity extent of the 12 absorption components (1–12), highlighting coincident absorption across multiple doublets and \lya. The SDSS \siiv\ spectrum is included solely for qualitative profile comparison. Minor wavelength offsets arise from instrumental differences across the observing setups, and all quantitative analyses presented in this work use only the X‑shooter datasets.}\label{fig:J0046}
\end{figure*}

Our analysis focuses on the \civ, \nv, and \siiv\ doublets, along with \lya. The normalized spectra in velocity space, together with all identified absorption components, are presented in \Cref{fig:J0046_total,fig:J0046}. We modeled each absorption component following the procedure described by \citet{Chen19}. Candidate \civ, \siiv, \nv, and \lya\ NALs were fitted with Voigt optical-depth profiles,
\begin{equation}
\label{eq:1}
\begin{split}
\tau(v) & =\tau_0\,H(a,u),\qquad \tau_0=\frac{\sqrt{\pi}\,e^2}{m_e c}\,\frac{f\lambda_0 N}{b},\\
u & =\frac{v-v_0}{b},\qquad a=\frac{\Gamma\,\lambda_0}{4\pi\,b},
\end{split}
\end{equation}
where $H(a,u)$ is the Voigt function, $\tau_0$ is the line-center optical depth, $v_0$ is the centroid velocity, $b$ is the Doppler parameter, $f$ and $\Gamma$ are the oscillator strength and damping constant of the transition, $\lambda_0$ is its laboratory wavelength, and $N$ is the ionic column density of the component. For doublet transitions, the two members were fitted simultaneously, with a common velocity centroid and Doppler parameter.

To account for partial coverage of the background emission source, we adopted the standard homogeneous partial-covering model \citep{Ganguly99, Hamann97b, Barlow97b}, which assumes a spatially uniform emission source and an absorber with the same optical depth along all covered sightlines. Under these assumptions, the normalized residual intensity is
\begin{equation}
\label{eq:2}
\frac{I_{v}}{I_{0}}=1-C_v+C_ve^{-\tau_{v}},
\end{equation}
where $I_0$ and $I_v$ are the continuum and observed intensities, respectively, and $C_v$ is the covering fraction of the absorber, constrained to $0<C_v\leq1$. Following previous studies of narrow associated absorption lines, we assume that the covering fraction remains constant across each absorption profile, such that $C_v=C_0$. This approximation is generally adequate for the relatively narrow features analyzed in this work.

Before comparison with the data, the model profiles were convolved with a Gaussian instrumental line-spread function with a velocity dispersion of $\sigma_{\rm inst}=23.6$~\kms, corresponding to an effective resolving power of $R_{\rm eff}\approx5400$ for the co-added UVB spectrum (Section~2.1). The best-fit parameters were obtained by least-squares minimization of the residuals between the convolved model and the normalized spectrum, and the quoted $1\sigma$ uncertainties were derived from the covariance matrix at the best-fit solution. For the weakest components, the covering fraction was fixed at $C_0=1$ to break the degeneracy among $C_0$, $N$, and $b$ \citep{Barlow97b, Hamann97b}.

\section{Results}
\subsection{Line Fitting Parameters}

\begin{figure*}[ht!]
\centering
\includegraphics[width=0.9\textwidth]{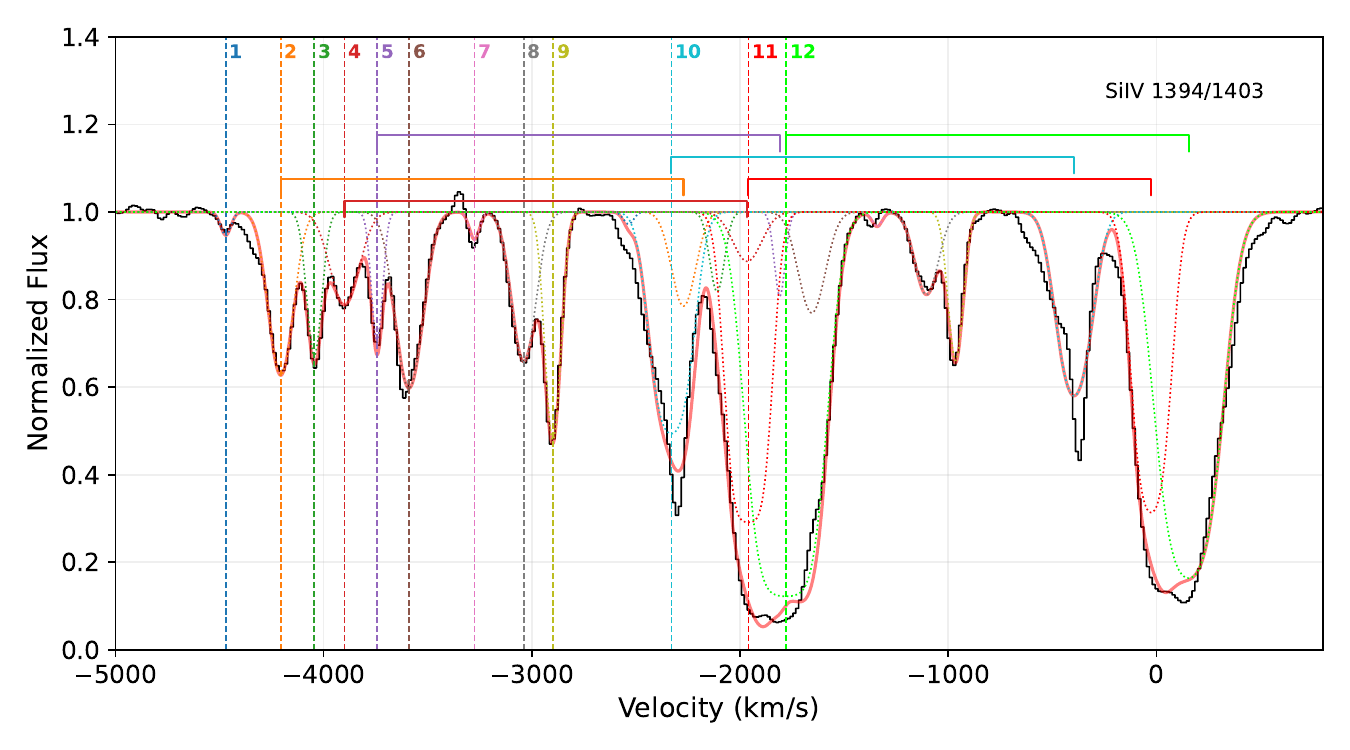}
\caption{Normalized \siiv\ line profiles in the X-shooter spectra plotted on a velocity scale relative to the quasar redshift $z=2.150$. The spectra are shown in black, and best-fit profiles in red. The dashed lines are identified components, and the brackets show the line-locked doublets. The velocities pertain to the short-wavelength lines in the doublets.\label{fig:siiv}}
\end{figure*}

\begin{figure*}[ht!]
\centering
\includegraphics[width=0.9\textwidth]{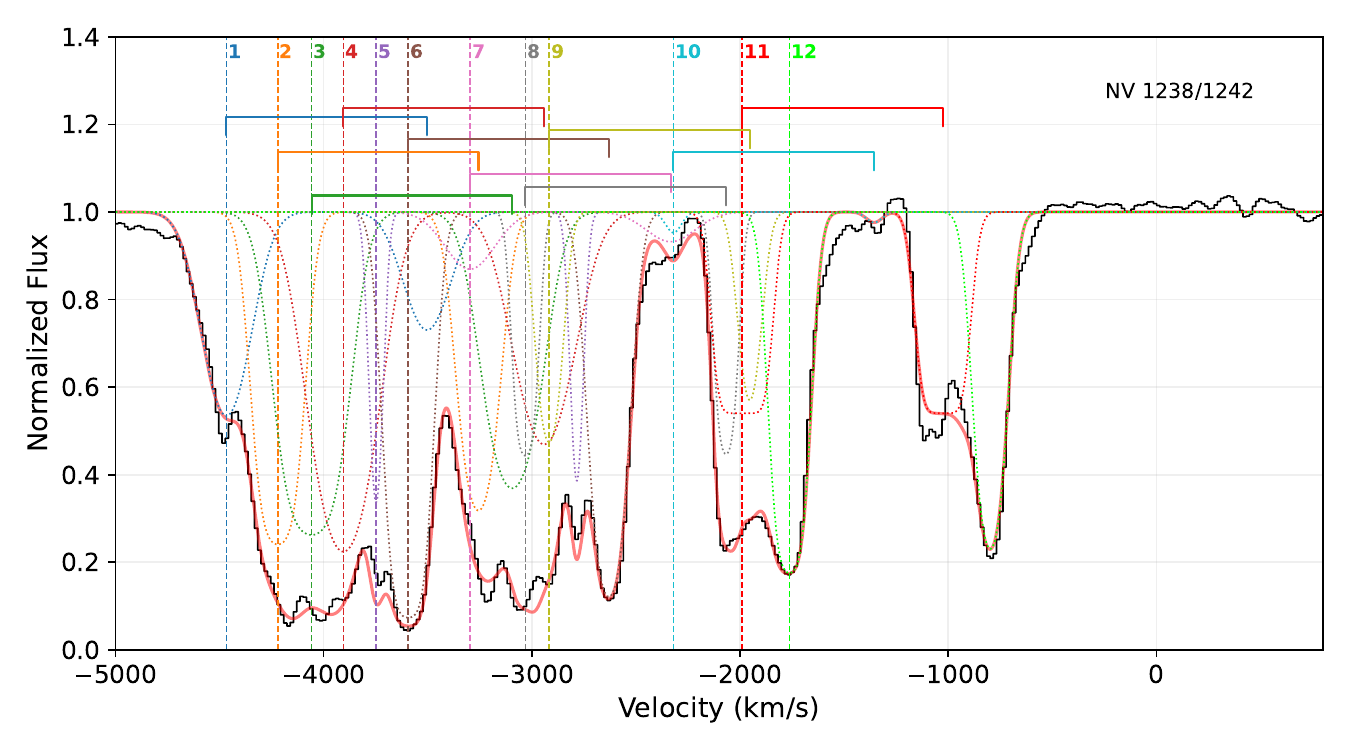}
\caption{Normalized \nv\ line profiles in the X-shooter spectra plotted on a velocity scale relative to the quasar redshift $z=2.150$. The spectra are shown in black, and best-fit profiles in red. The dashed lines are identified components, and the brackets show the line-locked doublets. The velocities pertain to the short-wavelength lines in the doublets.\label{fig:nv}}
\end{figure*}

\begin{figure*}[ht!]
\centering
\includegraphics[width=0.9\textwidth]{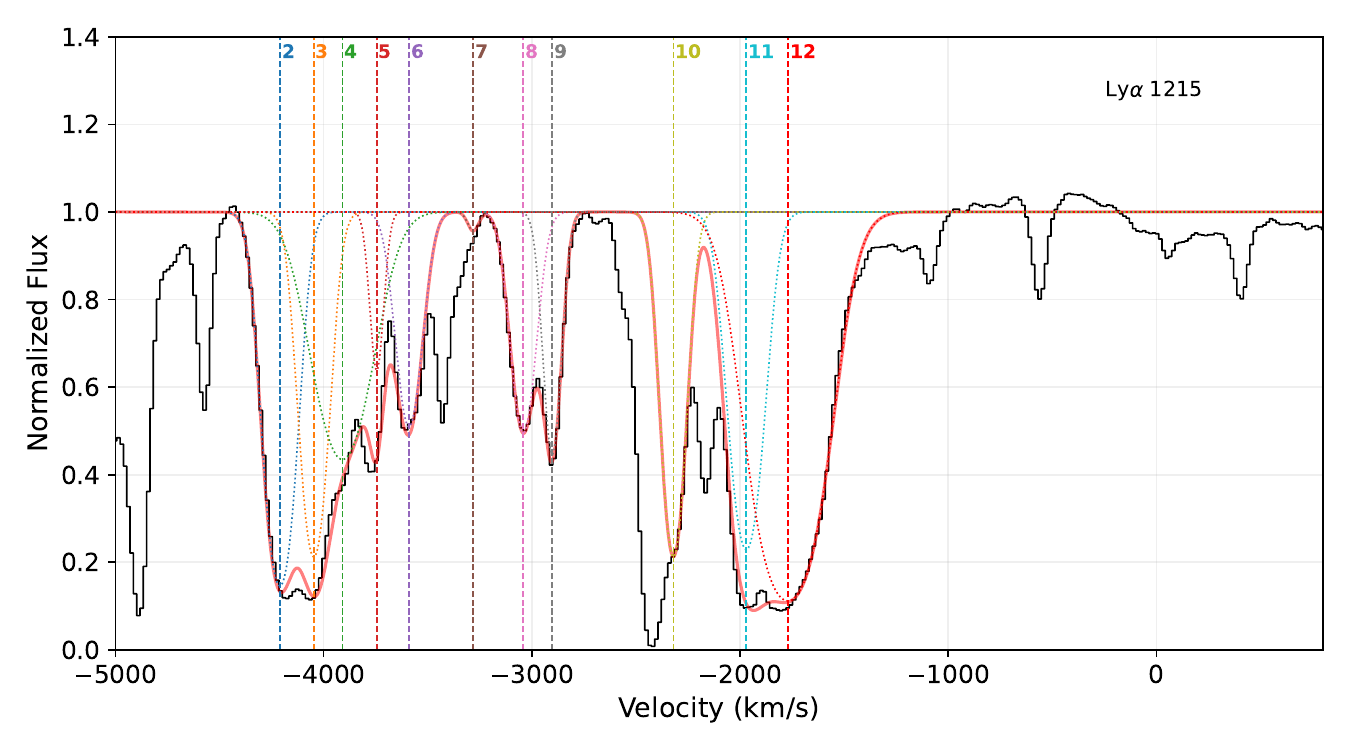}
\caption{Normalized \lya\ line profiles in the X-shooter spectra plotted on a velocity scale relative to the quasar redshift $z=2.150$. The spectra are shown in black, and best-fit profiles in red. \label{fig:lya}}
\end{figure*}

The fitted parameters and associated uncertainties for the \civ, \siiv, \nv, and \lya\ NALs identified in the X-shooter spectrum of quasar \target\ are listed in \Cref{tab:J0046}. For each absorption component, the table reports the velocity shift $v$, line identification, rest-frame wavelength, observed wavelength, Doppler parameter $b$, logarithmic column density, and covering fraction $C_0$. Measurements for the \civ, \siiv, \nv, and \lya\ transitions are presented separately. The absorbers are organized by redshift and component number, corresponding to the labeling adopted in \Cref{fig:J0046,fig:siiv,fig:nv,fig:lya}.

For most components, the quoted uncertainties correspond to the $1\sigma$ errors returned by the profile-fitting procedure and are dominated by the pixel-to-pixel noise in the spectra. For severely blended features, parameter values and limits were estimated following the methods described in Section 3.2 of \citet{chen18} and Section 3.1 of \citet{Chen19}.

The profile fits for all cataloged \siiv\ doublets are shown in \Cref{fig:siiv}. The velocity shift $v$, covering fraction $C_0$, and column density $\log N$ were derived by fitting the \siiv\ doublet profiles using Equations~(\ref{eq:1})--(\ref{eq:2}). Corresponding fits for the \nv\ and \lya\ associated absorption lines are presented in \Cref{fig:nv} and \Cref{fig:lya}, respectively. Given the substantial blending of the \civ\ doublet components, we do not carry out full line-profile fitting for this ion. We place lower limits on the \civ\ column densities using Doppler parameters constrained from the \siiv\ doublets.

\begin{table*}
\caption{Individual absorption lines of \target. Columns show component number, velocity shift ($v$) relative to the emission-line redshift 2.150, line identification and rest wavelength, Doppler b parameter, logarithm of column density, covering fraction, and notes.}
\label{tab:J0046}
\centering
\renewcommand{\arraystretch}{1.2}
\begin{tabular}{ccccccc}
\hline\hline
$\#$ & $v$ (\kms) & ID & $b$ (\kms) & $\log N$ (\cmN) & $C_0$ & Notes\\
\hline
1 & -4470 & \lya\ \lam1215 & -- & -- & -- & no obvious line\\
 &  & \civ\ \lam1548,1551 & $83.5\pm33.0$ & $14.17\pm1.00$ & 1.0 & \\
 & -4467 & \nv\ \lam1239,1243 & $146.4\pm26.0$ & $14.51\pm1.02$ & 1.0 & \\ 
 & -4470 & \siiv\ \lam1394,1403 & $10.0\pm5.0$ & $12.27\pm0.78$ & 1.0 &  weak line\\
\hline
2 & -4210 & \lya\ \lam1215 & $85.1\pm23.0$ & $14.37\pm0.13$ & 1.0 & \\
 &  & \civ\ \lam1548,1551 & -- & $\gtrsim14.9$ & -- & adopt $b$ from \siiv\ \\
 & -4220 & \nv\ \lam1239,1243 & $98.0\pm9.2$ & $15.19\pm0.87$ & $0.77\pm0.50$ & \\ 
 & -4206 & \siiv\ \lam1394,1403 & $70.5\pm9.9$ & $13.64\pm0.20$ & $0.83\pm0.13$ & \\
\hline
3 & -4047 & \lya\ \lam1215 & $77.1\pm28.7$ & $14.24\pm0.14$ & 1.0 & \\
 &  & \civ\ \lam1548,1551 & -- & $\gtrsim14.7$ & -- & adopt $b$ from \siiv\ \\
 & -4058 & \nv\ \lam1239,1243 & $149.3\pm41.5$ & $15.27\pm0.51$ & $0.76\pm0.16$ & \\ 
 & -4045 & \siiv\ \lam1394,1403 & $36.0\pm11.5$ & $13.27\pm0.45$ & $1.00\pm0.23$ & \\
\hline
4 & -3910 & \lya\ \lam1215 & $178.4\pm53.2$ & $14.30\pm0.35$ & 1.0 & \\
 &  & \civ\ \lam1548,1551 & -- & $\gtrsim15.0$ & -- & adopt $b$ from \siiv\ \\
 & -3905 & \nv\ \lam1239,1243 & $178.2\pm55.6$ & $14.98\pm2.02$ & $0.99\pm0.35$ & \\ 
 & -3900 & \siiv\ \lam1394,1403 & $88.7\pm44.8$ & $13.33\pm0.81$ & $0.99\pm0.45$ & \\ 
\hline
5 & -3745 & \lya\ \lam1215 & $30.8\pm17.8$ & $13.45\pm0.26$ & 1.0 & \\
 &  & \civ\ \lam1548,1551 & -- & $\gtrsim14.3$ & -- & adopt $b$ from \siiv\ \\
 & -3748 & \nv\ \lam1239,1243 & $19.9\pm10.0$ & $15.00\pm1.20$ & $0.78\pm0.45$ & \\ 
 & -3744 & \siiv\ \lam1394,1403 & $14.7\pm27.8$ & $13.39\pm1.47$ & $0.72\pm0.50$ & \\ 
\hline
6 & -3590 & \lya\ \lam1215 & $75.9\pm15.5$ & $13.87\pm0.13$ & 1.0 & \\
 &  & \civ\ \lam1548,1551 & -- & $\gtrsim14.8$ & -- & adopt $b$ from \siiv\ \\
 & -3594 & \nv\ \lam1239,1243 & $86.0\pm14.0$ & $15.27\pm0.38$ & $0.93\pm0.10$ & \\ 
 & -3590 & \siiv\ \lam1394,1403 & $80.2\pm10.1$ & $13.67\pm0.15$ & $0.92\pm0.12$ & \\ 
\hline
7 & -3282 & \lya\ \lam1215 & $14.7\pm12.1$ & $12.34\pm0.12$ & 1.0 & \\
 &  & \civ\ \lam1548,1551 & -- & $\gtrsim14.3$ & -- & adopt $b$ from \siiv\ \\
 & -3297 & \nv\ \lam1239,1243 & $159.9\pm91.4$ & $13.92\pm2.58$ & $0.95\pm0.48$ & \\ 
 & -3276 & \siiv\ \lam1394,1403 & $7.1\pm20.5$ & $12.35\pm1.15$ & 1.0 & weak line\\ 
\hline
8 & -3041 & \lya\ \lam1215 & $72.3\pm11.8$ & $13.87\pm0.04$ & 1.0 & \\
 &  & \civ\ \lam1548,1551 & -- & $\gtrsim15.1$ & -- & adopt $b$ from \siiv\ \\
 & -3031 & \nv\ \lam1239,1243 & $32.6\pm20.5$ & $15.80\pm2.06$ & $0.56\pm0.55$ & \\ 
 & -3039 & \siiv\ \lam1394,1403 & $69.7\pm7.2$ & $13.48\pm0.40$ & $1.00\pm0.52$ & \\ 
\hline
9 & -2902 & \lya\ \lam1215 & $47.0\pm4.5$ & $13.80\pm0.05$ & 1.0 & \\
 &  & \civ\ \lam1548,1551 & -- & $\gtrsim14.7$ & -- & adopt $b$ from \siiv\ \\
 & -2917 & \nv\ \lam1239,1243 & $56.4\pm19.1$ & $14.83\pm0.37$ & $0.56\pm0.22$ & \\ 
 & -2899 & \siiv\ \lam1394,1403 & $36.1\pm4.0$ & $13.72\pm0.13$ & $0.80\pm0.14$ & \\ 
\hline
10 & -2320 & \lya\ \lam1215 & $64.5\pm4.1$ & $14.18\pm0.06$ & 1.0 & \\
 &  & \civ\ \lam1548,1551 & -- & $\gtrsim15.0$ & -- & adopt $b$ from \siiv\ \\
 & -2320 & \nv\ \lam1239,1243 & $41.4\pm32.1$ & $12.95\pm0.39$ & 1.0 & weak line\\ 
 & -2330 & \siiv\ \lam1394,1403 & $89.6\pm4.7$ & $14.45\pm0.06$ & $0.53\pm0.03$ & \\ 
\hline
11 & -1970 & \lya\ \lam1215 & $93.9\pm14.8$ & $14.34\pm0.12$ & 1.0 & \\
 &  & \civ\ \lam1548,1551 & -- & $\gtrsim15.0$ & -- & adopt $b$ from \siiv\ \\
 & -1990 & \nv\ \lam1239,1243 & $75.1\pm49.8$ & $16.00\pm1.53$ & $0.46\pm0.10$ & \\ 
 & -1960 & \siiv\ \lam1394,1403 & $77.2\pm9.0$ & $14.74\pm0.13$ & $0.71\pm0.05$ & \\ 
\hline
12 & -1770 & \lya\ \lam1215 & $196.1\pm23.3$ & $14.76\pm0.11$ & 1.0 & \\
 & -1770 & \civ\ \lam1548,1551 & $161.2\pm30.5$ & $15.80\pm0.59$ & $0.87\pm0.05$ & \\
 & -1762 & \nv\ \lam1239,1243 & $73.1\pm11.5$ & $15.17\pm0.77$ & $0.83\pm0.27$ & \\ 
 & -1780 & \siiv\ \lam1394,1403 & $137.6\pm6.0$ & $14.90\pm0.05$ & $0.88\pm0.02$ & \\ 
\hline
\end{tabular}
\end{table*}

\begin{figure*}[ht!]
\centering
\includegraphics[width=0.95\textwidth]{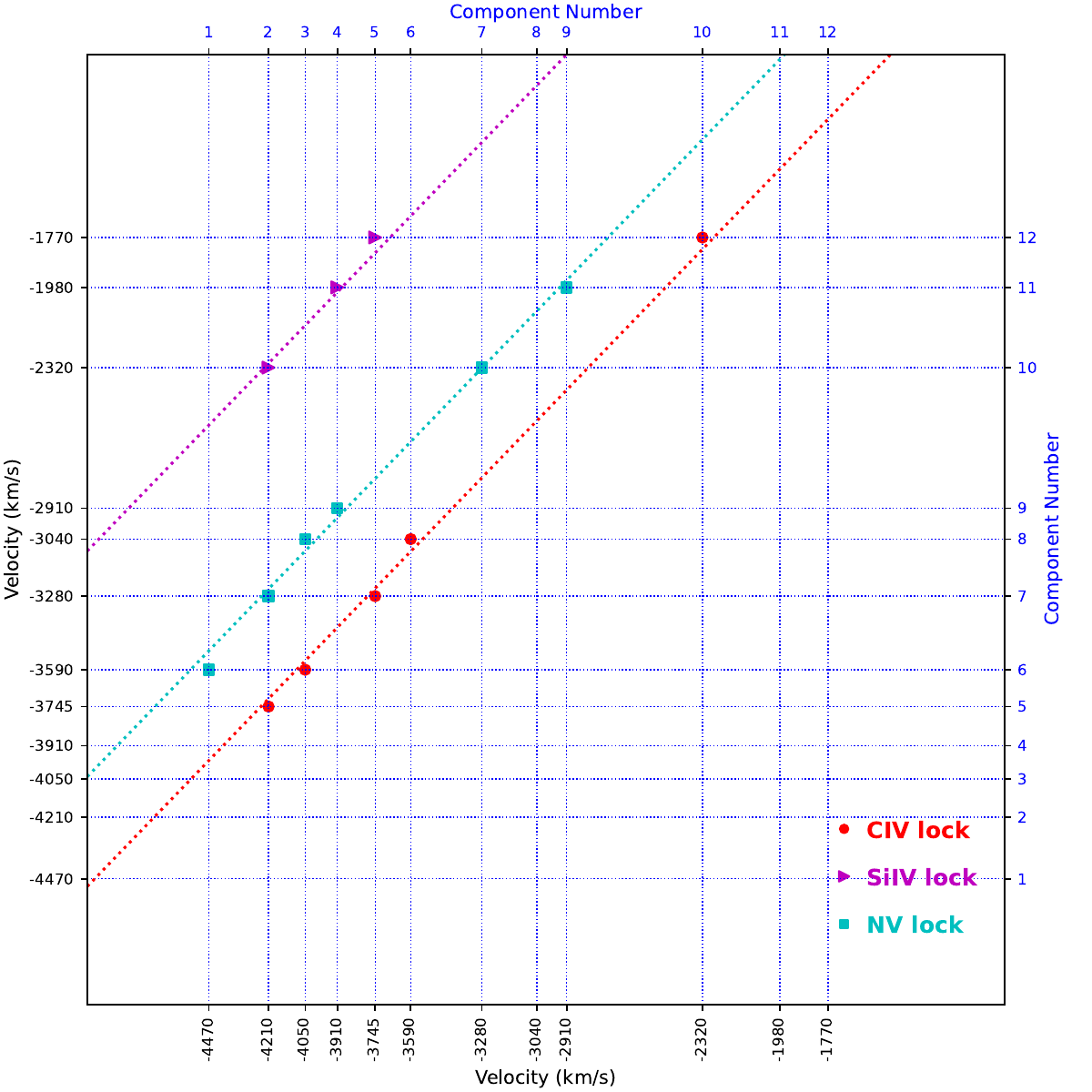}
\caption{The component matrix shows all possible line locks in \civ, \siiv, and \nv\ based on their velocity separations. We consider two systems locked when their velocity separation is a close match (with errors less than 10\%) to their laboratory doublet separation (see text for details). All line-locking pairs corresponding to the same ion doublet sit on the same line. The red line marks the \civ\ doublet, the purple line \siiv, and the cyan line \nv.}
\label{fig:matrix}
\end{figure*}

\subsection{Line-locked Signatures \label{sec:lock_result}}
We searched for line-locked absorption systems associated with the \civ, \siiv, and \nv\ doublets in the spectrum of quasar \target. The search was performed using the velocity-space component matrix shown in \Cref{fig:matrix}. Within this representation, line-locked absorbers appear along straight lines with a slope of $+1$, where the intercept corresponds to the velocity splitting of the relevant ionic doublet. The characteristic velocity separations are approximately $498~\rm{km~s^{-1}}$ for \civ, $964~\rm{km~s^{-1}}$ for \nv, and $1939~\rm{km~s^{-1}}$ for \siiv.

Two absorbers are classified as line-locked when their velocity separation satisfies
\begin{equation}
\mid{\delta v-v_{\rm sep}} \mid \le \min(\rm{b_1}, \rm{b_2}, 0.1\times v_{\rm sep}),
\end{equation}
where $v_{\rm sep}$ denotes the velocity separation of the corresponding doublet, and $\rm{b_1}$ and $\rm{b_2}$ are the Doppler parameters of the two absorbers. Confirmed line-locked pairs are marked by colored brackets in \Cref{fig:siiv,fig:nv}. 

In \Cref{fig:siiv}, three line-locked pairs of \siiv\ absorbers are identified: pairs (2, 10), (4, 11), and (5, 12) with separations matching the \siiv\ \lam1393, 1403 doublet. We also detect line-locking among other ionic doublets, including 6 pairs in \nv\ \lam1239, 1243, and 5 pairs in \civ\ \lam1548,1551. For \nv\, 4 line-locked groups are found: (1, 6), (2, 7, 10), (3, 8), and (4, 9, 11). In the case of \civ\, 3 line-locked NAL groups are identified: (2, 5, 7), (3, 6, 8), and (10, 12). 

\begin{figure*}[ht!]
\centering
\includegraphics[width=0.95\textwidth]{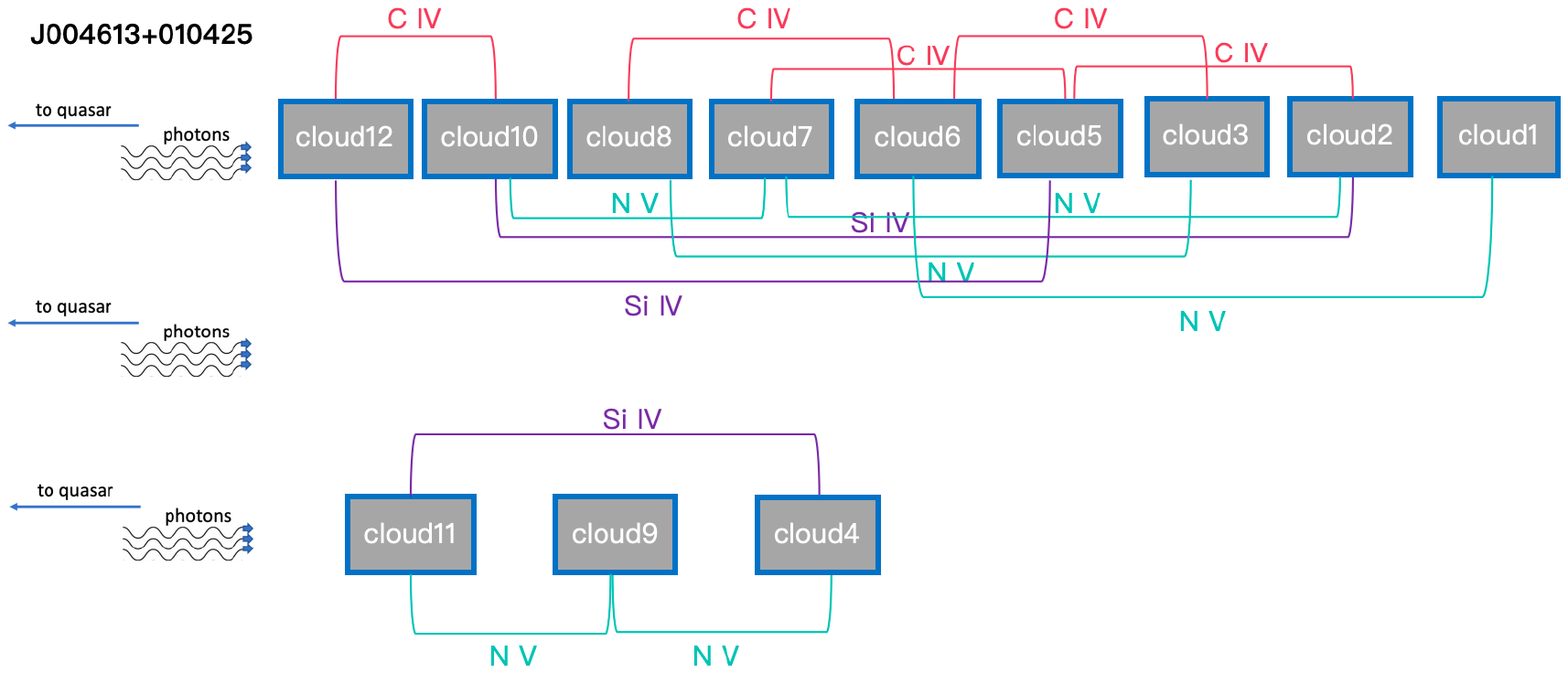}
\caption{Schematic illustration of the complex absorber systems line-locked through multi-ion doublets detected in the spectrum of \target. All absorbers are exposed to ionizing radiation coming from the quasar on the left. 
The number inside the cloud box is the corresponding component index shown in \Cref{tab:J0046}. 
}
\label{fig:multi-lock}
\end{figure*}

The velocity-space component matrix (\Cref{fig:matrix}) reveals a complex network of absorbers connected through velocity separations consistent with the doublet splittings of \civ, \siiv, and \nv. Such interconnected structures, often referred to as line-locking webs \citep{Chen25}, arise when multiple absorption components satisfy line-locking conditions associated with different ionic transitions. In total, twelve absorption components participate in the web identified in \target. A schematic illustration of the network is shown in \Cref{fig:multi-lock}.

Such web configuration could occur by chance. To assess the likelihood of a chance alignment, we followed the procedure of \citet{Srianand02}. We randomly distributed 12 absorbers over the velocity interval from $-5000$ to $-1000~\rm{km~s^{-1}}$ and searched for configurations matching that shown in \Cref{fig:multi-lock}. 
We repeated the experiment $10^{5}$ times and found no matching configurations, corresponding to a chance probability of $<10^{-5}$.

The presence of multiple line-locked doublets within the same absorption complex, together with the extremely low probability of a random alignment, indicates that the observed velocity structure is unlikely to arise by chance. Similar line-locking configurations have previously been reported in other quasar outflows \citep[e.g.,][]{Ganguly08, Hamann11, Chen25}. The absorber network identified here therefore provides an additional example of the close connection between line-locking phenomena and radiatively accelerated outflows \citep{Srianand02, Juranova24}.

Theoretical investigations of line-locking have traditionally relied on simplified two-cloud models \citep{Lewis23}. The identification of an extensive multi-component line-locking web, however, highlights the need for more sophisticated modeling. Exploring potential interactions among clouds within such networks represents a promising direction for future research.

\subsection{Photoionization model \label{sec:photo}}
To constrain the physical conditions of the absorbers, we performed photoionization modeling using \cloudy\ (version 2023; \citealt{Chatzikos23}). This approach allows us to estimate the ionization states, total hydrogen column densities, and elemental abundances of the NAL components. Each of the twelve components identified in Section~\ref{sec:lock_result} has measured column densities for at least two ionic species (\Cref{tab:J0046}), providing multiple observational constraints.

\begin{figure*}[ht!]
\centering
\includegraphics[width=0.95\textwidth]{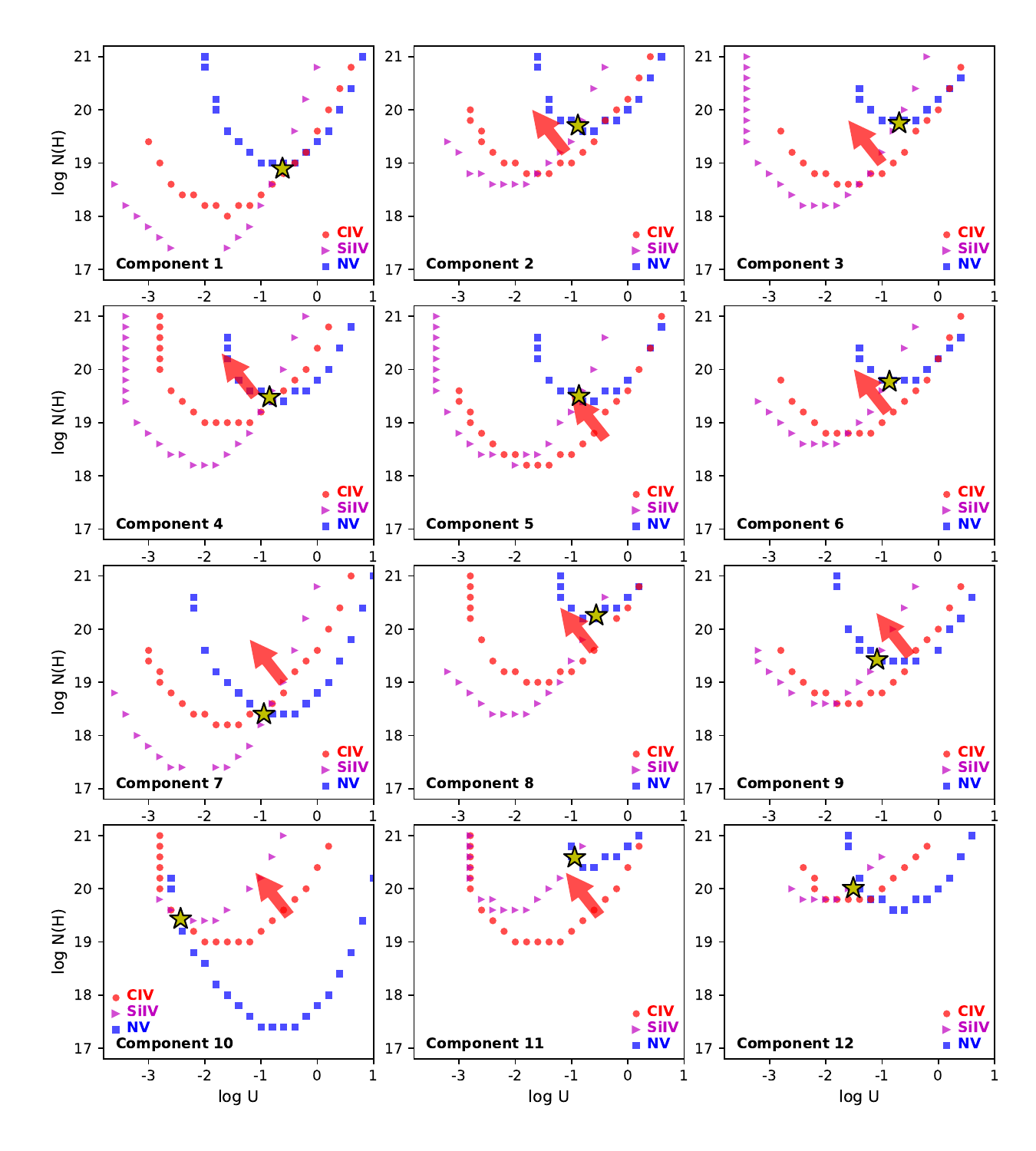}
\caption{Results from the Cloudy model for components 1-12 on the log $U$-log $N$(H) grid assuming solar metallicity. The red, purple and blue symbols mark the results from simulating the column densities of \civ\, \siiv\, and \nv, respectively. The area where different colored symbols meet shows the most likely model parameters.}
\label{fig:cloudy}
\end{figure*}

In the \cloudy\ model, we assume the gas is in photoionization equilibrium under the incident ionizing spectral energy distribution (SED) and adopt a solar metallicity for all calculations. The incident ionizing continuum is the Cloudy built-in AGN SED \citep{Mathews87, Chatzikos23}, implemented with the \texttt{AGN} command and parameterized by a Big Bump (blackbody) temperature $T=2\times10^{5}$~K together with the spectral indices $\alpha_{\rm ox}=-1.6$, $\alpha_{\rm uv}=-0.5$, and $\alpha_{\rm x}=-0.9$.
We construct a two-dimensional parameter grid spanning the ionization parameter 
$\log U$ and total hydrogen column density $\log N$(H), with uniform grid steps of 0.1 dex along both axes.
The model results from \cloudy\ are shown in \Cref{fig:cloudy}, from which we constrain the total hydrogen column density $\log N$(H) and the ionization parameter $\log U$ for components 1-12. 
The resulting values span $\log N(\rm{H})~(\rm{cm}^{-2}) \sim 18.4$ to $20.6$ and $\log U \sim -2.4$ to $-0.6$.

We then estimated metallicities using the relation:
\begin{equation}
\left[\frac{\rm{M}}{\rm{H}}\right] = \log \left( \frac{N(\rm{M}_i)}{N(\hi)} \right) + \log \left( \frac{f(\hi)}{f(\rm{M}i)} \right) + \log \left( \frac{\rm{H}}{\rm{M}} \right){\odot},
\end{equation}
where $(\rm{H/M})_{\odot}$ is the solar hydrogen-to-metal ratio, $N(\hi)$ and $f(\hi)$ are the column density and ionization fraction of neutral hydrogen, and $N(\rm{M}_i)$ and $f(\rm{M}_i)$ are the corresponding quantities for a given ion of element M \citep{chen18, Chen19}. Ionization fractions were obtained from the \cloudy\ models for the derived $U$ values, and measured column densities were used to compute the metallicities [M/H].

\begin{figure*}[ht!]
\centering
\includegraphics[width=0.95\textwidth]{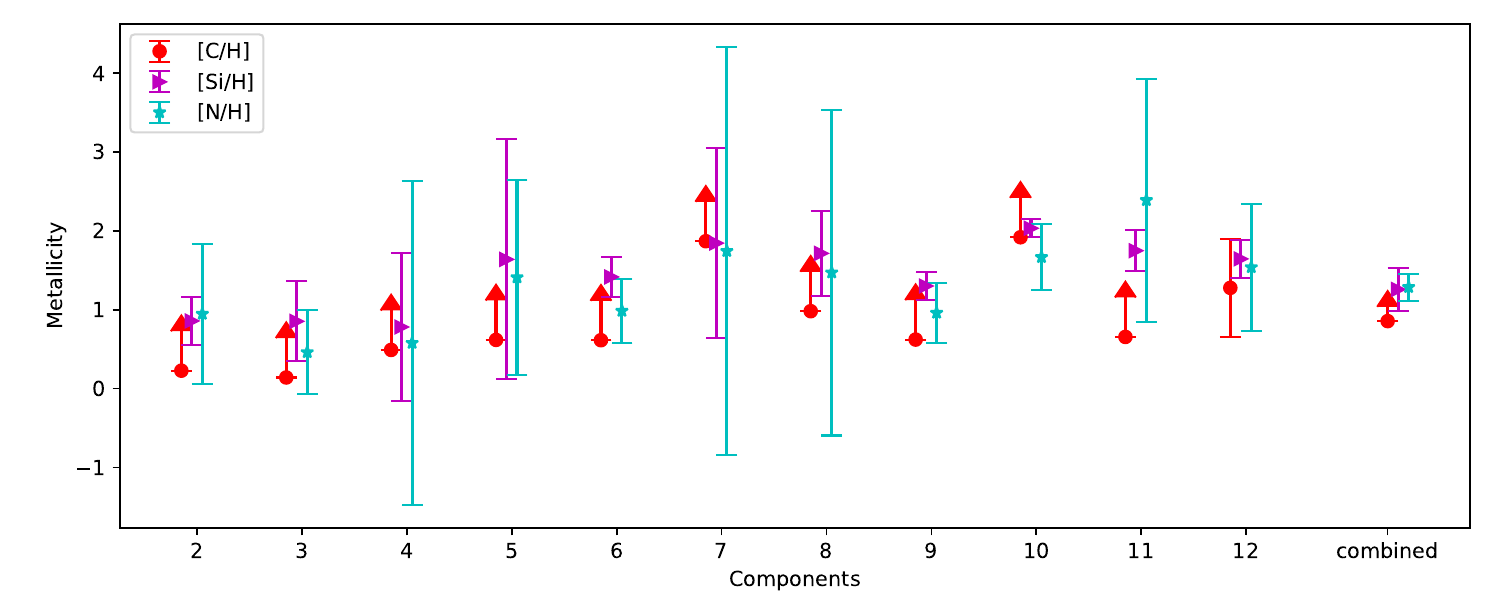}
\caption{C, Si, N abundances for components 2-12.}
\label{fig:metallicity}
\end{figure*}

The derived abundances for Si, N, and C are presented in \Cref{fig:metallicity}, with error bars reflecting measurement uncertainties in the ionic column densities. Metallicity estimates are available for components 2 to 12, which have reliable $N(\hi)$ measurements. For metallicity estimates of C, we only have lower limits for most of the components because the lines are blended and saturated (see \Cref{fig:J0046}). 
The combined metallicities for Si, N, and C are $\sim1.3$, $\sim1.3$, and $\gtrsim0.8$, respectively, indicating that the absorption components are metal-rich, which may facilitate the formation of such an extended line-locking web.
Previously, \citet{Fechner09} found 27$\%$ of the associated systems show clearly enhanced nitrogen \citep[see also][]{Petitjean94,Petitjean08}. In Papers I and II, we found that 20$\%$ (1 out of 5) of the NAL systems show evidence of enhanced nitrogen. Here for \target\, it appears that most components show evidence of [N/C]~$>0$. However, we caution that the estimated C abundances are only lower limits. Thus this result should be considered preliminary and warrants further investigation.

Assuming ten times solar metallicity, cloud sizes ranging from 0.1 to 1~pc, and hydrogen column densities of $\log N(\textrm{H})$(\cmN) ranging from 18.4 to 20.6 as derived from \cloudy, we estimate the masses of typical absorbers to lie between $1 \times 10^{-3}$ and $3$~$M_\odot$.
Following the method of Paper I, we estimate the total kinetic energy of the outflow system identified through line-locking, and evaluate its impact on the host galaxy's evolution from an energetic standpoint. 
Assuming typical distances of $\sim$10–100~pc for the line-locking clouds, we obtain kinetic energies $K \lesssim 1-100 \times 10^{52}$~ergs, and total kinetic luminosity of $\dot{K} \lesssim (0.1-1) \times 10^{42}$~ergs~s$^{-1}$ for this line-locking web. The resulting ratio between the kinetic energy rate and bolometric luminosity, $\dot{K}/L_{\rm Bol} \lesssim (0.2-1.6) \times 10^{-4}$, is about two orders of magnitude lower than the threshold of 0.005 to 0.05 required for effective feedback on the host galaxy \citep{Scannapieco04, DiMatteo05, Prochaska09, Hopkins10}. 
%logLbol=45.8
This result is consistent with previous studies of NAL outflow systems \citep{chen18, Chen19, Chen24, Chen25}.

\section{Discussion}
\subsection{Comparison with previous studies}\label{sec:comparison}

Quasar \target\ is one of only four known quasars exhibiting LL signatures in \civ, \siiv, and \nv\ doublets simultaneously. The other three are Q1303+308 \citep{Turnshek84a, Foltz87, Braun89}, J221531$-$174408 (Paper I), and SDSS~J092345+512710 \citep{Lin20b}. As illustrated in Figure~\ref{fig:matrix}, \target\ exhibits one of the most complex LL structures reported in the literature. All twelve identified NAL components are linked through radiative forces, as evidenced by their velocity separations corresponding to the \civ, \siiv, and \nv\ doublets. This is further consistent with our previous statistical study based on BOSS (Baryon Oscillation Spectroscopic Survey) survey data, which found that intrinsic NALs generally have velocities lower than $\sim8000$ km/s \citep{Chen20a}. 

\citet{Lu25} first identified LL phenomena in \target\ using seven epochs of SDSS spectroscopy obtained between 2010 and 2020. Our analysis employs co-added VLT/X-shooter observations obtained in 2012--2013 at significantly higher spectral resolution. Below we present a systematic comparison of the two studies across four key aspects: data quality, component identification, line-locking architecture, and physical modeling.

The most fundamental difference between the two studies lies in the spectral resolution of the data employed. \citet{Lu25} utilized SDSS spectra with $R \sim 2000$, corresponding to a velocity resolution of $\sim 150$~\kms\ at the observed wavelengths of the \siiv\ and \civ\ doublets. In contrast, our X-shooter spectra achieve $R \approx 3251$--$8935$ (see Table~\ref{tab:data}), yielding velocity resolutions of $\sim 35$--$90$~\kms\ across the relevant wavelength range. 
All the transitions analyzed here (\civ, \nv, \siiv, and \lya) fall in the X-shooter UVB arm, whose nominal resolving power is $R\sim5400$ ($\sigma_{\rm inst}\approx23.6$~\kms) for a 1.0'' slit  in the line-profile fits.
This improvement in resolving power by a factor of $\sim 2.7$ is critical for line-locking analyses, where the ability to separate closely spaced NAL components directly determines the completeness of the detected LL network.

The higher spectral resolution of X-shooter also enables more accurate measurements of line centroid velocities and Doppler parameters. At SDSS resolution, NAL components separated by less than $\sim 200$~\kms\ are typically blended into a single feature, whereas our X-shooter data can resolve components with separations as small as $\sim 50$--$100$~\kms. This difference has direct implications for the identification of LL pairs, as the velocity separation criterion ($|\delta v - v_{\rm sep}| \leq \min(b_1, b_2, 0.1 \times v_{\rm sep})$; see Section~\ref{sec:lock_result}) depends sensitively on the accuracy of the measured component velocities.

\citet{Lu25} reported five \siiv\ NAL doublets within BAL system B of \target, which they used as the primary basis for their LL analysis. Our higher-resolution X-shooter spectra reveal a substantially more complex absorption structure: we identify twelve distinct NAL components (Table~\ref{tab:J0046}) through the \siiv, \civ, \nv, and \lya\ transitions. The seven additional components detected in our work are either fully blended within the lower-resolution SDSS profiles or too weak to be distinguished from the noise at $R \sim 2000$.

The increased component count has significant consequences for the inferred outflow geometry. While \citet{Lu25} interpreted the absorption system as comprising five relatively homogeneous clouds, our analysis suggests a more fragmented outflow structure with twelve distinct absorbing elements spanning a velocity range of $\sim 2700$~\kms\ (from $v \approx -4470$ to $-1770$~\kms). This finer granularity is consistent with the picture emerging from other high-resolution studies of quasar outflows \citep[e.g.,][]{Arav18, Xu19, Miller20}, which similarly reveal complex substructures within BAL troughs that are invisible in low-resolution spectra.

The differences in component identification propagate directly into the detected LL architecture. \citet{Lu25} identified LL signatures associated primarily with the \civ\ and \nv\ doublet splittings. Our analysis confirms the presence of LL pairs corresponding to these doublets but additionally reveals several important differences:

\begin{enumerate}
    \item {\bf \siiv\ line-locking:} We identify three LL pairs with separations matching the \siiv\ doublet ($\Delta v \approx 1939$~\kms): pairs (2, 10), (4, 11), and (5, 12). These pairs were not reported by \citet{Lu25}, likely because the \siiv\ doublet lines are partially blended at SDSS resolution, making it difficult to assign unambiguous component velocities for LL matching. The detection of \siiv\ LL pairs is particularly significant because \siiv\ has the largest doublet splitting among the commonly observed UV resonance lines, and LL signatures in \siiv\ provide strong constraints on the radiative acceleration mechanism \citep{Scargle73, Goldreich76}.

    \item {\bf \nv\ line-locking:} We detect six LL pairs corresponding to the \nv\ doublet ($\Delta v \approx 964$~\kms), organized into four LL groups: (1, 6), (2, 7, 10), (3, 8), and (4, 9, 11). These results are broadly consistent with the \nv\ LL signatures reported by \citet{Lu25}, although our higher-resolution data allow us to resolve individual components within the groups that appear merged in the SDSS spectra.

    \item {\bf \civ\ line-locking:} We identify five LL pairs corresponding to the \civ\ doublet ($\Delta v \approx 498$~\kms), organized into three groups: (2, 5, 7), (3, 6, 8), and (10, 12). \citet{Lu25} reported two \civ\ LL pairs; however, one of their reported pairs (the b--c pair in their notation) is considered marginal in our analysis due to poor profile matching between the two components at high resolution. The discrepancies in \civ\ LL identification arise primarily from the severe blending of the \civ\ doublet in \target, which makes component decomposition highly sensitive to spectral resolution and fitting methodology.

    \item{\bf \aliii\ line-locking:} \citet{Lu25} reported LL signatures associated with the \aliii\ doublet ($\Delta v \approx 1302$~\kms). We do not detect corresponding \aliii\ LL pairs in our X-shooter data (\Cref{fig:J0046}). Although weak \aliii\ absorption is present in components 10 and 12 individually, their velocity difference is inconsistent with the line-locking condition.

\end{enumerate}

In total, the LL network identified in this work comprises fourteen distinct LL pairs across three ionic species, connecting all twelve NAL components into a single line-locking web (Figure~\ref{fig:matrix}). This represents a significant expansion over the LL architecture reported by \citet{Lu25}, which consisted of approximately seven LL pairs across \civ, \nv, \siiv, and \aliii. The probability of the observed twelve-component LL configuration arising by chance is $< 10^{-5}$ (Section~\ref{sec:lock_result}), strongly supporting a physical origin via radiative acceleration.

\citet{Lu25} did not carry out photoionization modeling in their work. The physical parameters of their single absorber candidate were originally taken from \citet{Xu19}. We compare our Cloudy-derived results for component 12 (matching their component 5) with the values published in \citet{Xu19}, and find consistent agreement within mutual measurement uncertainties. Specifically, they reported a neutral hydrogen column density $\log N(\hi)~({\rm cm}^{-2}) \gtrsim 15.6$, an ionization parameter $\log U \sim -1.4$, and a total hydrogen column density $\log N({\rm H})~({\rm cm}^{-2}) \sim 21.2$. Our \cloudy\ modeling for component~12 yields $\log N(\hi)~({\rm cm}^{-2}) \sim 14.8$, $\log U \sim -1.5$, and $\log N({\rm H})~({\rm cm}^{-2}) \sim 20.0$. These differences likely reflect the different spectral resolutions of the input data, the limited signal-to-noise ratio of the \lya\ absorption measurement, and the associated uncertainties in the measured column densities.

A major advantage of our higher-resolution data is the ability to perform photoionization modeling for all twelve components rather than a single representative absorber. As summarized in Table~\ref{tab:J0046} and Figures~\ref{fig:cloudy}--\ref{fig:metallicity}, we derive ionization parameters ranging from $\log U \approx -2.4$ to $-0.6$ and total hydrogen column densities spanning $\log N({\rm H})~({\rm cm}^{-2}) \sim 18.4$--$20.6$. The full set of component parameters enables us to construct a substantially more complete picture of the outflow ionization structure, including the detection of significant metallicity variations among the locked components (Figure~\ref{fig:metallicity}). \citet{Lu25} did not report metallicity measurements, precluding a direct comparison on this point.

The systematic comparison demonstrates that low-resolution spectroscopy, while valuable for identifying candidate LL systems in large quasar samples, is inherently limited in its ability to fully resolve the complex substructure of line-locked outflows. Medium- and high-resolution observations are essential for (1) detecting faint and closely spaced NAL components that are blended at low resolution; (2) accurately measuring component velocities for robust LL identification; (3) resolving ambiguities in LL pair assignments arising from component blending; and (4) performing comprehensive photoionization modeling of all absorbers in a multi-component LL network. These findings reinforce the conclusions of Papers I and II that medium- and high-resolution spectroscopy is indispensable for detailed studies of quasar line-locking phenomena.

\subsection{Spectra variability analysis}

\citet{Lu25} analyzed seven epochs of SDSS spectroscopy of \target\ and found significant variability in the absorption systems over a multi-year baseline. The observed changes were correlated with continuum variations, suggesting that fluctuations in the ionization state of the absorbing gas play an important role in driving the spectral variability.

We compared our X-shooter spectra obtained between November 2012 and August 2013 with the SDSS spectra analyzed by \citet{Lu25}. As shown in Figure~\ref{fig:J0046_total}, no significant differences are apparent among the X-shooter spectra obtained at adjacent epochs. The absorption profiles, including their depths and velocity structures, remain consistent within the measurement uncertainties. This lack of short-term variability contrasts with the larger changes observed across the SDSS epochs.
The simplest interpretation is that the absorber was in a comparatively stable ionization state during the X-shooter observing campaign. If the variability reported by \citet{Lu25} is primarily governed by changes in the ionizing continuum, the absence of detectable spectral changes may indicate that the continuum remained relatively stable during this period. 

The absence of measurable variability over the $\sim0.7$ yr observed-frame interval ($\sim0.2$ yr in the quasar rest frame) suggests that the characteristic ionization-variation timescale of the absorber is longer than the temporal baseline sampled by the X-shooter observations. This is consistent with the picture proposed by \citet{Lu25}, in which substantial changes in the absorption systems occur primarily on multi-year timescales. The X-shooter observations therefore likely captured the quasar during a relatively quiescent phase, when continuum variations were insufficient to produce detectable changes in the ionic absorption lines. Combining the SDSS and X-shooter observations, the available data suggest that the absorber remains stable on timescales shorter than $\sim1$ yr, while significant ionization-driven variability can develop over intervals of several years.

\subsection{`Shadowed' vs `shadowing' absorbers}\label{sec:shadow}

\begin{figure}[htbp]
\centering
\includegraphics[width=0.48\textwidth]{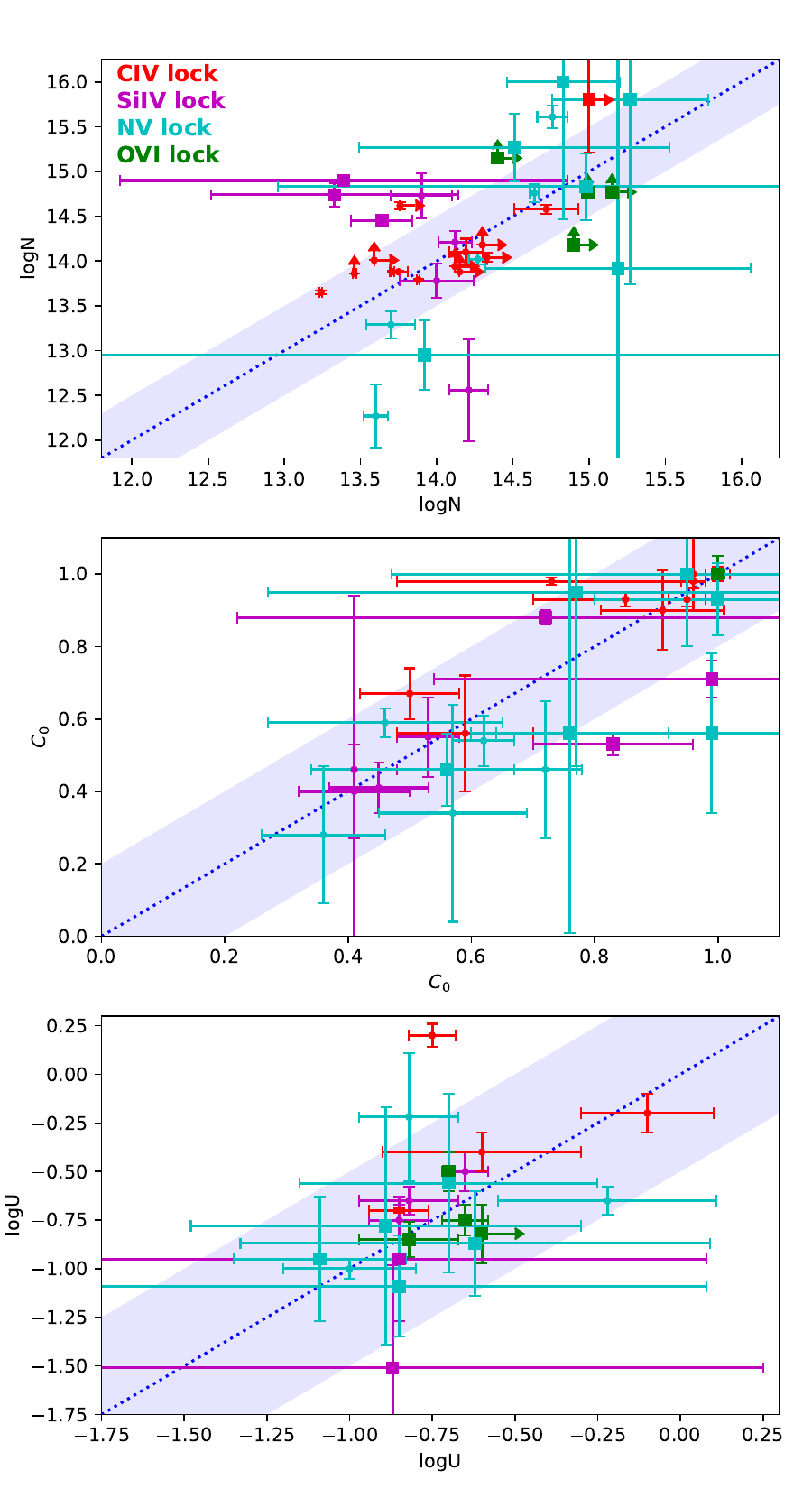}
\caption{Comparison between the derived properties of the shadowed cloud (x-axis) and shadowing cloud (y-axis) in the same line-lock pair. 
Upper panel: logarithm of column density. Middle panel: covering fraction. Lower panel: logarithm of ionization.
The blue dashed line represents the 1:1 ratio. In the upper and lower panels, the blue shaded regions delineate the region where the parameters of the shadowing clouds are within a factor of $\sim3$ above and below those of the shadowed clouds. In the middle panel, the blue shadow marks the 20\%\ uncertainty area around the 1:1 ratio line.
The sample contains all the line-locked systems that were analyzed in \citet{chen18,Chen19,Chen24,Chen25} (represented by filled circles) and this paper (represented by the filled squares).}
\label{fig:N_U_C}
\end{figure}

All twelve NAL components identified in \target\ participate in line-locking pairs, a locking fraction of $\sim$100\% that is notably higher than the $\sim$65\% reported for the general quasar population with multiple \civ\ NALs at velocities $\lesssim 12,000$~\kms\ \citep{Bowler14}. This high locking fraction, together with the large-scale LL web architecture presented in Section~\ref{sec:lock_result}, indicates that radiative acceleration plays a dominant role in organizing the velocity structure of the \target\ outflow \citep{Milne26, Scargle73, Braun89, Lewis23, Dannen24}.

In the standard line-locking picture, a slower ``shadowing'' cloud shields a faster ``shadowed'' cloud from specific resonance-line photons, producing a stable velocity separation equal to the relevant doublet splitting \citep{Goldreich76}. Under this scenario, the two clouds are expected to have comparable physical properties, since the radiation field that drives the locking must also govern the ionization balance of both absorbers. We tested this prediction by comparing the ion column densities, covering fractions ($C_0$), and ionization parameters ($\log U$) between shadowed and shadowing clouds in all LL pairs identified in \target, supplemented by the LL systems analyzed in \citet{chen18}, \citet{Chen19}, \citet{Chen24}, and \citet{Chen25}. The results are shown in Figure~\ref{fig:N_U_C}.

We find that the ion column densities and ionization parameters of shadowing clouds typically lie within a factor of $\sim$3 of those of the corresponding shadowed clouds, and their covering fractions are mostly consistent within $\sim$20\%. 
We carried out Pearson linear, Spearman rank and $\tau$-b correlation analyses for three physical quantities linking shadowed clouds and shadowing clouds: logarithmic column density ($\log N$), covering fraction, and ionization parameter ($\log U$). For $\log N$, all three tests yield statistically significant coefficients ($r=0.5143$, $p=0.0013$; $\rho=0.5437$, $p=0.0006$; $\tau=0.3943$, $p=0.0008$), indicating a moderate positive correlation. For covering fraction, all three metrics show extremely significant results ($r=0.7975$, $\rho=0.8129$, $\tau=0.6417$, all $p<0.0001$), indicating a strong positive correlation. For $\log U$, the parametric Pearson test returns a weak, non-significant linear correlation ($r=0.3707$, $p=0.0745$), yet the two non-parametric tests produce highly significant coefficients ($\rho=0.6193$, $p=0.0013$; $\tau=0.4541$, $p=0.0025$), robustly revealing a moderate monotonic positive association rather than a strict linear relationship between the ionization parameters of the two cloud populations.
These trends broadly support the simulation results of \citet{Lewis23}, who argued that line-locked clouds require similar physical conditions to maintain stable velocity separations. However, we also identify several outliers in which the shadowing and shadowed clouds exhibit column density ratios exceeding 10 or covering fractions differing by more than a factor of 2. Such exceptions challenge the simplified two-cloud, single-doublet model adopted by \citet{Lewis23} and suggest that additional physical processes such as multi-ion locking, cloud--cloud interactions, or time-dependent ionization effects may regulate the stability of LL configurations in complex outflows. 
The relatively poor correlations may also result from the limited spectral resolution of the present X-shooter data, which prevents the internal structure of the absorption systems from being fully resolved and introduces significant uncertainties into the inferred cloud parameters (column densities, covering fractions, and ionization parameters).
A quantitative assessment of these possibilities will require future simulations incorporating $\gtrsim 10$ clouds and $\geq 3$ ionic doublets simultaneously, which is beyond the scope of this paper.

\subsection{Physical implications of line locking}\label{sec:ll_physics}

Statistical studies find that line locking occurs in a substantial fraction of multi-NAL systems \citep{Bowler14}, while the TOLL project has revealed increasingly complex examples involving multiple ions and multiple absorbing components using high-quality spectral observations (Papers I and II). These observations provide several insights into the structure and acceleration of quasar outflows.

First, line locking provides evidence that radiative forces on resonance lines can influence the velocity structure of the absorbing gas. When two absorbing components have a velocity separation close to the splitting of a strong resonance doublet, the downstream component can be preferentially shielded from the radiation that accelerates the upstream component, allowing radiative forces to couple their velocities \citep{Milne26, Scargle73, Burbidge75, Goldreich76, Foltz87, Braun89, Lewis23}. The high incidence of line locking in multi-NAL systems ($\sim$65\% in the statistical study of \citealt{Bowler14}) suggests that this mechanism may operate over a substantial range of outflow conditions. The present system provides another particularly clear example, with all twelve identified components participating in at least one line-locking relationship.

Second, previous studies show that line locking is not limited to simple pairs of clouds or to a single ionic species. For example, line locking was identified simultaneously through C~IV, N~V, and Si~IV, in a large locking web involving multiple absorbers (Papers I and II). The present system adds another example of this complex behavior. These multi-ion locking structures suggest that radiative acceleration, ionization, and shielding are coupled across different phases of the outflow. In this sense, line locking provides an observational link between the velocity and ionization structures of the absorbing gas and offers a way to study how clouds with different physical conditions are organized within the flow.

Third, the physical properties of the locked components provide constraints on the conditions under which line locking can occur. In particular, the similarity between the physical conditions of the shadowed and shadowing clouds in the present system is broadly consistent with the radiative-acceleration picture in which the relative shielding of neighboring clouds regulates their acceleration \citep{Lewis23}. At the same time, the existence of outliers and the diversity of the physical properties inferred in different TOLL targets indicate that a simple two-cloud picture is unlikely to describe the full complexity of quasar outflows. Multi-ion coupling, interactions among multiple clouds, and time-dependent ionization may all play a role and should be considered in more realistic models of line-locked flows.

The properties of the gas also provide clues to the broader environment in which line locking occurs. The metal-rich composition inferred for the locked components in \target\ is consistent with efficient line driving, since a larger abundance of metals provides a richer set of UV resonance transitions through which radiation can transfer momentum to the gas. This suggests a possible connection between the chemical enrichment of the nuclear environment and the efficiency of radiative acceleration. However, the different metallicities inferred for individual TOLL targets indicate that line locking does not require a single characteristic chemical composition, and a larger sample will be needed to determine whether the metallicity of the absorbing gas systematically affects the occurrence or strength of line locking.

An important next step is to determine whether line-locking structures remain stable over time. Multi-epoch spectroscopy can distinguish between a transient velocity coincidence and a dynamically maintained line-locked configuration. If the velocity separations between locked components remain close to the corresponding ionic doublet splittings while the line strengths, covering fractions, or column densities vary, this would provide stronger evidence that the velocity relationship is dynamically maintained rather than arising from a chance alignment. Conversely, changes in the locked velocity separations or the disappearance of previously identified locking relationships would constrain the lifetime and stability of the locked structures, as well as their response to variations in the ionizing continuum. Multi-epoch monitoring can therefore provide a direct probe of the timescale on which radiative forces establish and maintain velocity locking, complementing the detailed physical constraints obtained from single-epoch spectroscopy. The long-term monitoring of Q1303+308 provides an early example in which line-locking signatures remained identifiable despite substantial changes in the absorption strengths over several years in the quasar rest frame \citep{Vilkoviskij01}. Extending such monitoring to TOLL targets would provide a valuable time-domain test of the stability and evolution of line-locked outflows.

The relatively modest kinetic luminosities inferred for the individual locked clouds ($\dot{K}/L_{\rm Bol}\lesssim10^{-4}$) suggest that systems such as \target\ are primarily valuable as laboratories for the physics of outflow acceleration rather than as major contributors to the overall AGN feedback budget. The broader significance of studies similar to the TOLL project is therefore to use resolved line-locking structures as a microscopic probe of quasar winds: from the velocity relationships between individual clouds, to their ionization and chemical properties, and ultimately to their temporal evolution.

\section{Conclusions}
We have presented a detailed investigation of LL signatures in the quasar \target\ ($z_{\rm em}=2.150$) using medium-resolution VLT/X-shooter spectroscopy from the TOLL project. The associated absorption complex is resolved into twelve narrow absorption-line (NAL) components traced in \siiv, \civ, \nv, and \lya\ and measured through Voigt-profile fitting with an explicit instrumental line-spread function, increasing the component count by a factor of $\sim 2.4$ compared to previous SDSS-based studies. We identify fourteen distinct LL pairs across three ionic species—three in \siiv, six in \nv, and five in \civ. All twelve components participate in line-locking, providing strong statistical evidence that radiative acceleration plays a dominant role in organizing the velocity structure of this outflow.

A direct comparison with the lower-resolution SDSS analysis of \citet{Lu25} demonstrates that spectral resolution critically affects LL identification. At $R\sim2000$, blending and limited sensitivity caused seven of the twelve components to be missed or ambiguously identified, and several LL pairs, particularly those in \siiv, remained unresolved due to doublet blending. Photoionization modeling with \cloudy\ yields ionization parameters $\log U \approx -2.4$ to $-0.6$ and total hydrogen column densities $\log N(\mathrm{H}) \approx 18.4$–$20.6$ for all components. While shadowed and shadowing clouds generally show consistency within a factor of $\sim 3$, a subset of systems exhibits column density ratios exceeding 10, suggesting that additional processes such as multi-ion coupling and cloud–cloud interactions may contribute to LL stability beyond the simplest radiative-locking picture. No significant variability is detected over the $\sim0.7$ yr baseline, consistent with longer ionization-response timescales inferred in \citet{Lu25}.

Overall, the twelve-component LL web in \target\ represents one of the most complex line-locking structures observed to date, extending beyond the canonical two-cloud, single-doublet framework commonly assumed in theoretical models. These results highlight the necessity of medium- and high-resolution spectroscopy for fully characterizing quasar outflow dynamics and for revealing the true multiplicity of LL configurations. Future expansion of the TOLL sample, combined with simulations incorporating multi-cloud, multi-doublet radiative coupling, will enable robust statistical constraints on LL occurrence and provide deeper insight into the role of radiative forces in shaping quasar winds and AGN feedback.

\section*{acknowledgments}

We thank the anonymous referee for a careful reading of the manuscript and for constructive comments that helped improve the paper. This work was supported in part by the National Natural Science Foundation of China (12103097). C.C. acknowledges the support of the Tri-level Talents Development Project of Zhuhai College of Science and Technology. W.-J.L. and Y.-R.L. acknowledge the support of the Guangxi Natural Science Foundation (2025GXNSFAA069632) and the National Natural Science Foundation of China (12463003). Based on observations collected at the European Organisation for Astronomical Research in the Southern Hemisphere under ESO programs 090.B-0577(A) and 091.B-0324(B).

\bibliographystyle{aa}
\bibliography{reference}

\end{document}